\documentclass[11pt,a4paper,twocolumn]{article}
\usepackage[utf8]{inputenc}
\usepackage[T1]{fontenc}
\usepackage[english]{babel}
\usepackage{amsmath, bm}
\usepackage{graphicx}
\usepackage{authblk}
\usepackage[scale=0.85]{geometry}
\usepackage[charter]{mathdesign}
\usepackage{caption}
\usepackage{booktabs}
\usepackage{array}
\usepackage{ragged2e}
\usepackage{multirow}
\usepackage{dcolumn}
\usepackage[colorlinks=true, citecolor=blue, linkcolor=blue, urlcolor=blue]{hyperref}
\usepackage{pgfplots,pgfplotstable}
\usepackage{xurl}
\pgfplotsset{width=8.5cm,compat=1.9}
\usetikzlibrary{angles,calc,positioning}
\usetikzlibrary{plotmarks}

\everymath{\displaystyle}
\DeclareMathOperator*{\argmin}{\arg\!\min}

\title{Nowcasting Madagascar's real GDP using machine learning algorithms}

\author[1]{Franck Ramaharo}
\author[2]{Gerzhino Rasolofomanana}

\affil[1]{\small Service de la Modélisation Économique, Ministère de l'Économie et des Finances}
\affil[2]{\small Service du Suivi des Indicateurs et de la Conjoncture, Ministère de l'Économie et des Finances\authorcr 
	Antananarivo 101, Madagascar\authorcr 
	\{\ttfamily$^1$\href{mailto:franck.ramaharo@gmail.com}{franck.ramaharo}, $^2$\href{mailto:mrgherme@gmail.com}{mrgherme}\}@gmail.com
}

\date{December 23, 2023}

\newcommand{\GDP}{\mathit{GDP}}
\newcommand{\ELEC}{\mathit{ELEC}}
\newcommand{\CARB}{\mathit{PETR}}
\newcommand{\FDI}{\mathit{FDI}}
\newcommand{\CUREX}{\mathit{GCUREX}}
\newcommand{\CAPEX}{\mathit{GCAPEX}}
\newcommand{\XG}{\mathit{XG}}
\newcommand{\MG}{\mathit{MG}}
\newcommand{\CRED}{\mathit{CRED}}
\newcommand{\TOUR}{\mathit{TOUR}}
\newcommand{\VAT}{\mathit{VAT}}
\everymath{\displaystyle}

\makeatletter
\newcolumntype{B}[3]{>{\boldmath\DC@{#1}{#2}{#3}}c<{\DC@end}}
\makeatother

\newcommand{\gras}[1]{\multicolumn{1}{B{.}{.}{5.2}}{#1}}
\newcommand{\grasi}[1]{\multicolumn{1}{B{.}{.}{4.2}}{#1}}
\newcommand{\grasii}[1]{\multicolumn{1}{B{.}{.}{4.2}}{#1}}
\newcommand{\grasiii}[1]{\multicolumn{1}{B{.}{.}{3.1}}{#1}}
\newcommand{\Figs}[1]{\hyperref[#1]{Figure~\ref*{#1}}}
\newcommand{\Tabs}[1]{\hyperref[#1]{Table~\ref*{#1}}}
\newcommand{\Scn}[2]{\hyperref[#1]{Scenario~#2}}

\begin{document}
	
	\maketitle
	
	\begin{abstract}
		We investigate the predictive power of different machine learning algorithms to nowcast Madagascar's gross domestic product (GDP). We trained popular regression models, including linear regularized regression (Ridge, Lasso, Elastic-net), dimensionality reduction model (principal component regression), k-nearest neighbors algorithm ($k$-NN regression), support vector regression (linear SVR), and tree-based ensemble models (Random forest and XGBoost regressions), on 10 Malagasy quarterly macroeconomic leading indicators over the period 2007Q1--2022Q4, and we used simple econometric models as a benchmark. We measured the nowcast accuracy of each model by calculating the root mean square error (RMSE), mean absolute error (MAE), and mean absolute percentage error (MAPE). Our findings reveal that the Ensemble Model, formed by aggregating individual predictions, consistently outperforms traditional econometric models. We conclude that machine learning models can deliver more accurate and timely nowcasts of Malagasy economic performance and provide policymakers with additional guidance for data-driven decision making.
		\bigskip
		
	\end{abstract}
	\vfill
	
	\noindent\textbf{Avertissement}. {Le contenu de la présente publication n'engage que ses auteurs. Chacune des opinions exprimées est personnelle et ne peut en aucun cas être considérée comme représentative des points de vue du Ministère de l'Économie et des Finances ou de tout autres entités mentionnées dans ce document de travail.}
	\vfill
	
	\noindent\textbf{Disclaimer}. {The opinions expressed in this working paper are the sole responsibility of the authors and do not reflect the views of the Ministry of Economy and Finance or any other mentioned entities.}
	
	\section{Introduction}
	In economic context, nowcasting refers to the ability to estimate current  Gross Domestic products (GDP) before data official release. This technique relies on a diverse set of high frequency indicators and other real-time economic variables to generate rapid and accurate estimates \cite{Evans2005,Giannoneetal2008}. It enables policymakers and researchers to gain insights into current economic conditions, especially in situations where official data may be incomplete or subject to delays. 
	
	Machine Learning algorithms have now become a valuable tool in economic modelling, demonstrating remarkable efficacy in the challenging task of nowcasting and forecasting GDP across diverse global contexts. This effectiveness is evident in advanced economies  (e.g., Canada \cite{Qureshietal2020}, China \cite{Xia2023, Zhangetal2023}, Finland \cite{FornaroLuomaranta2020}, Italy \cite{Ciccerietal2020}, Netherlands \cite{Kantetal2022}, New Zealand \cite{Richardsonetal2018,Richardsonetal2019,SusnjakSchumacher2018}, South Africa \cite{Bothaetal2021}, Sweden \cite{Jonsson2020}, USA \cite{Hopp2022, Maas2019}, multiple European countries \cite{Dauphinetal2022}), emerging markets and developing countries (e.g., Albania \cite{VikaKlea2022}, Bangladesh \cite{Hossainetal2021}, Belize and El Savador \cite{Barriosetal2021}, Brazil \cite{RossideOliveira2023}, Egypt \cite{AbdEl-Aal2023}, Georgia \cite{Mgebrishvili2022}, India \cite{GhoshRanjan2023}, Indonesia \cite{Tamaraetal2021}, Lebanon \cite{Tiffin2016}, Malaysia \cite{Jasnietal2022}, Peru \cite{TenorioPerez2023})). Moreover, Machine learning algorithms are also proved to be very competitive with respect to standard econometric methods. 
	
	\begin{table*}[!htb]
		\centering
		\caption{Data description.}
		\begin{tabular}{@{}p{0.35\textwidth}lp{0.2\textwidth}l@{}}
			\toprule
			\textbf{Variables} & \textbf{Notation} & \textbf{Units} & \textbf{Source} \\
			\midrule
			Real gross domestic products & $\GDP$ & Billions of Ariary, at 2007 constant  prices & INSTAT \cite{INSTATCN2022b} \\
			Electricity consumption & $\ELEC$ & GWh & JIRAMA \& INSTAT \cite{INSTATCN2022c,JIRAMA2023}\\
			Petroleum consumption & $\CARB$ & Cubic decameter & OMH \& INSTAT \cite{INSTATCN2022c, OMH2021,OMH2023} \\
			Value-added tax & $\VAT$ & Billions of Ariary & Trésor Public Malagasy \cite{Tresor2023} \\
			Foreign direct investment & $\FDI$ & Millions of SDRs & BFM \cite{BFM2023a,BFM2023c}\\
			Credit to the economy & $\CRED$ & Billions of Ariary & BFM \cite{BFM2023a,BFM2023c}\\
			Government current expenditure (excl.\ debt interests payment) & $\CUREX$  & Billions of Ariary & Trésor Public Malagasy \cite{Tresor2023} \\
			Government capital expenditure & $\CAPEX$  & Billions of Ariary & Trésor Public Malagasy \cite{Tresor2023} \\
			Tourist arrivals & $\TOUR$ & Number of arrivals & MINTOUR \& INSTAT  \cite{INSTATCN2022c,MINTOUR2023} \\
			Exports of goods & $\XG$  & Kilotons & Douane Malagasy \cite{Douane2023}\\
			Imports of goods & $\MG$  & Kilotons & Douane Malagasy \cite{Douane2023}\\
			\bottomrule
		\end{tabular}%
		\label{tab:data}%
	\end{table*}%
	
	For Madagascar particularly, the task of nowcasting is more challenging due to the scarcity of high frequency indicators as well as their relatively short time span. In this study, we embrace this challenge by developing several machine learning models tailored to nowcast Madagascar's real GDP. We take into account the variability in model performance and we adopt the forecast combination technique. This strategy aims to mitigate the risks associated with relying solely on individual models, as our experiment revealed distinct performances among models in adapting to economic changes. 
	
	Our results show that in all cases, machine learning models provide smaller forecast errors than econometric models which are the workhorses of Malagasy economists. In fact, our study contributes to the ongoing dialogue on the adoption of advanced modeling techniques in economic forecasting, emphasizing the adaptability and superior performance of machine learning models in capturing the intricacies of Madagascar’s economic dynamics

	\section{Experimental setup}
	\subsection{Dataset and features}
	Our empirical assessment is grounded in the Keynesian theory of aggregate demand, which posits that changes in this demand determines the level of real output and employment in the short-run \cite{Blinder2008}. Hence, the proposed methodology relies on the aggregate demand equation:
	\begin{equation}\label{eq:Keynes}
		Y^{ad}_t = C_t + I_t + G_t + NX_t,
	\end{equation}
	where $Y^{ad}_t$ denotes aggregate demand, $C_t$ is private consumption, $I_t$ is private investment, $G_t$ is government expenditure, $NX_t$ are net exports. Within the context of Madagascar, the leading indicators listed in \Tabs{tab:data} are regarded as those capable of anticipating potential short-term fluctuations in the GDP.  Moreover, these indicators are consistent with existing empirical research on real-time monitoring of economic activities \cite{BentsenGorea2021, Bhattacharya2023, Dauphinetal2022, GulKazdal2021, KunovacSpalat2014, Proietti201, Prakashetal2023}. 
	
	Given the structure of  \eqref{eq:Keynes}, our primary model incorporates the following functional relationship:
	\begin{equation}\label{eq:generalform}
		\begin{split}
			\GDP_t = f(&\ELEC_t, \CARB_t, \VAT_t, \FDI_t, \CRED_t,\\
			&\CUREX_t,\CAPEX_t, \TOUR_t,  \XG, \MG_t),
		\end{split}
	\end{equation}
	where variables are defined in \Tabs{tab:data}. The dataset consist of  quarterly observations that cover the period 2007Q1 to 2022Q4, and are  issued by Malagasy authorities, including the National Institute of Statistics (INSTAT), the General Directorate of Customs (Douane Malagasy) and the General Directorate of Treasury (Trésor Public Malagasy) within the Ministry of Economy and Finance, the Malagasy Office of Hydrocarbons (OMH), Jiro sy Rano Malagasy (JIRAMA) -- the national electricity and water company of Madagascar, the Ministry of Tourism (MINTOUR), and the Central Bank of Madagascar (BFM).

	\subsection{Data prepocessing}
	We converted all nominal data into real data by deflating their value with the consumer price index \cite{INSTATIPC2022a}. Next we  applied \textbf{robust scaling} to the dataset which is done by removing the median and scaling to unit interquartile range, i.e.,  scaling the data in the range between the first quartile and third quartile. We used the formula	
	\begin{equation*}
		x_i^{\textit{scaled}} = \dfrac{x_i-Q_2(x)}{Q_3(x)-Q_1(x)},
	\end{equation*}
	where $Q_1$, $Q_2$ and $Q_3$ denote the first, the second (median) and the third quartile, respectively.

	\subsection{Data split}
	To conduct our analysis effectively, we divided our dataset into distinct training and test sets, by employing various partitions to assess the performance of the machine learning models across different economic scenarios. \Tabs{tab:scnTrainTest} illustrates the scenarios designed to evaluate model efficiency during four distinct economic periods:  a stable period (2019Q1--Q4) displaying a good economic performance compared to the last five years \cite{BFM2019}, a COVID-19 period (2020Q1--Q4) marked by an economic recession \cite{BFM2020,INSTAT20221}, a post-COVID-19 period (2021Q1--Q4) marked by an economic recovery \cite{BFM2021}, and a new situation (2022Q1--Q4) influenced by the Russian-Ukraine war \cite{Andrianadyetal2023,BFM2022}. The strategic division provides insights into the models' adaptability to varying economic conditions.	Note that we could have added an intermediate scenario between Scenarios 1 and 2, say Scenario 1$^{*}$, in which the dataset would be partitioned as 2007Q1--2019Q4 and 2020Q1--2022Q4 for the train and test set quarters, respectively. Because 2019 is a relatively stable year, our experiment showed that models trained in Scenario 1$^{*}$ performed as well as those trained in Scenario 1.
	
	\begin{table}[!htb]
		\centering
		\caption{Scenarios and dataset partition.}
		\begin{tabular}{@{}crr@{}}
			\toprule
			\textbf{Scenario} & Training set quarters & Test set quarters \\
			\midrule
			\textbf{1} & 2007Q1--2018Q4 & 2019Q1--2022Q4\\
			\textbf{2} & 2007Q1--2020Q4 & 2021Q1--2022Q4\\
			\textbf{3} & 2007Q1--2021Q4 & 2022Q1--2022Q4\\
			\bottomrule
		\end{tabular}%
		\label{tab:scnTrainTest}%
	\end{table}%

	\subsection{Machine learning algorithms for nowcasting exercises}
	We employ well-established machine learning methods that have demonstrated effectiveness in predicting economic aggregates \cite{Aguetal2022, Barhoumietal2022, Ciccerietal2020, Jungetal2018, NymanOrmerod2017, Qureshietal2020, Richardsonetal2019, VikaKlea2022}. The selected methods can be categorized into two groups:
	
	\begin{itemize}
		\item Parametric models: linear regularized regression (Ridge, Lasso, Elastic-net), dimensionality reduction model (principal component regression);
		\item Non-parametric models: $k$-nearest neighbors algorithm ($k$-NN regression), support vector regression (linear SVR), tree-based ensemble models (random forest regression, XGBoost regression).
	\end{itemize}
	
	For benchmarking purposes, we also use traditional econometric models, including a simple univariate autoregressive AR(4) model and multiple linear regression estimated by ordinary least squares. 
	
	In what follows,  we let the expression in \eqref{eq:lin} denote the linear estimate of \eqref{eq:generalform}: 
	\begin{equation}\label{eq:lin}
		\begin{split}
			y_t = \beta_0 + \beta_1 x_{1,t} + \beta_{2}x_{2,t} + \cdots + \beta_p x_{p,t} + \varepsilon_t,
		\end{split}
	\end{equation}
	where $\beta_i,\, i= 1,\ldots, p$ are unknown coefficient parameters of the predictors to be estimated, and $\varepsilon_t$ is the error term of the regression.

	\subsubsection{Autoregressive model}
	We used a univariate Autoregressive $AR(4)$ model as a simple benchmark for quarterly GDP growth. For convenience, the GDP variable is transformed into quarterly year-over-year growth in natural logarithmic form, i.e., $\widetilde{\GDP}_t = \log\GDP_t - \log\GDP_{t-4}$. The corresponding model is expressed as follows:
	\begin{equation*}
		\begin{split}
			\widetilde{\GDP}_t &= \phi_0 + \phi_1 \widetilde{\GDP}_{t-1}  + \phi_2 \widetilde{\GDP}_{t-2} + \phi_3 \widetilde{\GDP}_{t-3} \\
			&+ \phi_4 \widetilde{\GDP}_{t-4} + \varepsilon_t,
		\end{split}
	\end{equation*}
	where $\phi_0$ is a constant, $\phi_1,\phi_2,\phi_3$ and $\phi_4$  are parameters to be estimated, and $\varepsilon_t$ represents the usual residual term. While lacking the sophistication of machine learning models, the interpretability of the autoregressive model and its  established track record in capturing historical trends under simpler economic dynamics made it a valuable starting point for understanding the data and assessing the potential gains from advanced modeling techniques.

	\subsubsection{Ordinary least squares method}
	We additionally employed ordinary least squares (OLS) method to estimate the linear regression associated with \eqref{eq:generalform}. Recall that OLS method is an optimization process which involves finding the coefficient parameters that minimize the residual sum of squares, which is the sum of the squared differences between the observed and predicted values:
	\begin{equation*}\label{eq:OLS}
		\argmin\limits_{\beta_0, \beta_1, \ldots,\beta_p} \left\{\sum_{i=0}^{N}\left(y_i - \beta_0 - \sum_{j=1}^{p}\beta_j x_{j,i}\right)^2\right\}.
	\end{equation*}
	
	In classic econometrics, data is often log-transformed before running regressions. While this is common practice, we wanted to explore an alternative. We applied linear regression to both log-transformed and robustly scaled data. We'll call these models OLS-log and OLS-RS, respectively. This comparison will help us assess the effectiveness of robust scaling compared to the traditional log-transform.

	\subsubsection{Ridge, Lasso, Elastic-Net regression}
	To refine coefficient estimates and improve model robustness and prediction accuracy, we employ three distinct regularization techniques: Ridge penalizes large coefficients, Lasso (Least Absolute Shrinkage and Selection Operator) enforces sparsity by shrinking some to zero, and Elastic-Net (E-Net) combines both approaches for balanced shrinkage and selection.
	
	Their respective general forms are as follows:
	\begin{equation*}
		\begin{split}
			\textbf{Ridge}:      &\min\limits_{\beta_0, \beta_1, \ldots,\beta_p} \sum_{i=0}^{N}\left(y_i - \beta_0 - \sum_{j=1}^{p}\beta_j x_{j,i}\right)^2\\	
			&\mbox{subject to } \sum_{j=1}^{p} \beta_j^2 \leq s_R,\\
			\textbf{Lasso}:      &\min\limits_{\beta_0, \beta_1, \ldots,\beta_p} \sum_{i=0}^{N}\left(y_i - \beta_0 - \sum_{j=1}^{p}\beta_j x_{j,i}\right)^2\\	
			&\mbox{subject to } \sum_{j=1}^{p} |\beta_j| \leq s_L,\\   
			\textbf{E-Net}:      &\min\limits_{\beta_0, \beta_1, \ldots,\beta_p} \sum_{i=0}^{N}\left(y_i - \beta_0 - \sum_{j=1}^{p}\beta_j x_{j,i}\right)^2\\	
			&\mbox{subject to } (1-\alpha_E)\sum_{j=1}^{p} \beta_j^2 + \alpha_E \sum_{j=1}^{p}|\beta_j| \leq s_E,\\  
		\end{split}
	\end{equation*}
	where, $s_R$, $s_L$ and $s_E$ represent the shrinkage parameters associated with Ridge, Lasso, and Elastic-Net, respectively. The parameter $\alpha_E\in[0,1]$ is a weighting factor. These regularization terms help control the complexity of the models by penalizing the size of the coefficient parameters, and the choice of the $s$ parameters is a crucial tuning hyperparameter in these methods.

	\subsubsection{Principal component regression}
	Principal component regression (PCR) is a regression model that combines principal component analysis (PCA) with OLS methodology. The algorithm first calculates the principal components, denoted as $z_1, z_2, \ldots, z_p$, which are new uncorrelated variables obtained through linear combinations of predictors. PCR then applies a linear model by selecting $k\leq p$ principal components:
	\begin{equation*}\label{eq:PCR}
		y_t = \theta_0 + \theta_1 z_{1,t} + \cdots + \theta_p z_{k,t} + \varepsilon^Z_t.
	\end{equation*}
	This approach proves particularly valuable in the context of machine learning, where PCR's ability to reduce dimensionality through PCA while maintaining predictive accuracy showcases its utility in handling high-dimensional datasets and mitigating multicollinearity.

	\subsubsection{Support vector regression}
	Let  $\{(\bm{x}_1, y_1), \ldots, (\bm{x}_N, y_N)\} \subset \chi \times \mathbb{R}$ be a training data, where $\chi$ denotes the space of the input patterns (in our case $\chi=\mathbb{R}^p$).  Recall that an hyperplane is a linear function of the form:
	\begin{equation*}\label{eq:hyperplan}
		f(x) = \langle \bm{w},\bm{x}\rangle + b,
	\end{equation*}
	where $\langle\cdot,\cdot\rangle$ denotes the dot product, $\bm{w}$ is a weight vector in $\chi$ , $\bm{x}$ is a vector in $\mathbb{R}^p$, and $b$ defines the bias. Linear support vector regression (SVR) involves modeling the regression function with a hyperplane, as flat as possible, positioned at the center of a hyper-tube with a width of $2\mu$, containing the maximum training points. Additionally, it permits the inclusion of other values outside the hyper-tube, provided they fall within a predefined margin tolerance. This is a minimization problem, expressed as:
	
	\begin{equation*}
		\min \dfrac{1}{2}\Vert \bm{w}\Vert ^2 + C\sum_{i=1}^{N}\left(\xi_i+\xi_i^{\star}\right)
	\end{equation*}
	subject to the constraints
	\begin{equation*}
		\begin{split}
			y_i - \langle \bm{w},\bm{x}_i\rangle - b \leq \mu + \xi_i,\\
			\langle \bm{w},\bm{x}_i\rangle + b -y_i\leq \mu + \xi_i^{\star},\\
			\xi_i,\xi_i^{\star}\geq 0, i=1,\ldots,N,
		\end{split}
	\end{equation*}	
	where $\xi_i,\xi_i^{\star}$ correspond to errors above and below, and the constant $C>0$ is a hyperparameter that adjusts the trade-off between the allowed error and the flatness of the function $f$ \cite{SmolaScholkopf2004}. In practice, to find out the best-fitted function $f$, the optimal penalty factor $C$ and the thickness of the tube $\mu$ are chosen by minimizing the prediction accuracy calculated based on the cross-validation technique \cite{Ito2003}.

	\subsubsection{k-Nearest Neighbors regression}
	Regression using the k-Nearest Neighbors ($k$-NN) method is a prediction technique that relies on the observations (neighbors) closest to a given point to estimate its value, typically evaluated using Euclidean distance. An observation at quarter $t$ is represented by a data point whose components are $\bm{x}_t= (x_{1,t}, x_{2,t},\ldots,x_{p,t}) \in \mathbb{R}^p$, and is associated with an attribute $y_t$ which is the corresponding value of the GDP at that quarter. To predict the target attribute for a query point $\bm{x}_t'$, the algorithm averages the target attributes of its $k$ nearest neighbors. The choice of $k$ is crucial, because it determines the trade-off between noise reduction and over-smoothing in predictions.  
	
	\begin{figure*}[!htbp]
		\centering
		\caption{Illustration of the 5-fold forward chaining time-series cross validation.}
		\includegraphics[width=0.75\linewidth]{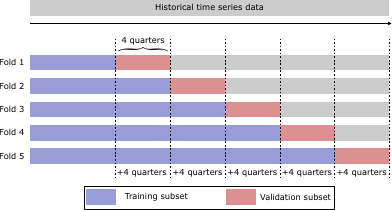}
		\label{fig:ts_crossval}
	\end{figure*}
	
	\subsubsection{Random forest regression}
	Random forest regression (RFR) is a powerful ensemble learning algorithm that combines multiple decision trees to make predictions in regression task \cite{Breiman2001}. It builds a large ``forest'' of decision trees during  the training phase, with each tree focusing on a random subset of features and data points. During the prediction phase, the individual tree predictions are averaged to obtain the final output.  This diverse ensemble architecture makes RFR particularly adept at capturing the complex, non-linear dynamics of economic data, leading to robust GDP predictions \cite{Barriosetal2021, GhoshRanjan2023, Jasnietal2022, Kantetal2022, Tamaraetal2021,Yoon2021}.

	\subsubsection{XGBoost}
	XGBoost Regression stands as a cutting-edge machine learning algorithm specifically designed for regression tasks, excelling in predicting  GDP \cite{ChuQureshi2023,KadhimReda2023,Mgebrishvili2022,Qureshietal2020,Xia2023}. It belongs to the gradient boosting family, a framework for sequentially combining weak learners (typically decision trees) to progressively improve prediction accuracy \cite{ChenGuestrin2016,Friedman2001}.
	
	At its core, XGBoost operates by iteratively minimizing a loss function, which quantifies the error between predicted and actual GDP values. In each iteration, it adds a new decision tree that focuses on correcting the largest remaining errors in the current ensemble. This adaptive learning process allows XGBoost to capture intricate, non-linear relationships within economic data, leading to significantly more accurate GDP predictions compared to simpler models.	Beyond its iterative optimization, XGBoost distinguishes itself by incorporating regularization terms. These penalize excessively large coefficients in the decision trees, effectively controlling model complexity and preventing overfitting.

	\subsubsection{Ensemble approach}
	After training the machine learning models, the ensemble approach is applied to consolidate their predictive capabilities. The Ensemble Model is constructed by computing a weighted mean of the individual model predictions \cite{BatesGranger1969,SmithWallis2009}. Each model's contribution to the final ensemble prediction is determined by its performance on the mean squared prediction error calculated on the test set. This ensemble approach combines the strengths of multiple models, and provides a balanced and robust prediction by assigning higher influence to well-performing models while mitigating the impact of models that may exhibit weaknesses.

	\subsection{Model selection}
	We used the cross-validation method for hyperparameters tuning. Since the data considered in this study are time series, the conventional cross-validation method is not suitable for model learning due to the temporal dependencies between data points and the arbitrary choice of test set implemented by these  methods. Therefore, we adopt forward chaining time-series cross validation \cite{BergmeirBenitez2012,Cochrane2018, Tashman2000} to perform model training and validation so that the best model can be chosen to generate the quarterly forecasts. In particular, we perform a {5-fold forward chaining time-series cross validation} which splits the training set into 5 training subsets, where the validation set consists of the 4 quarters immediately following the training subset. The first fold is used to train the model, which is then evaluated in the 4 quarters immediately following that fold. Afterward, the training subset increases by 4 values, and the validation subset is also moved 4 positions into the future, and the process is repeated until all of the folds have been exhausted (see \Figs{fig:ts_crossval}). The performance measure reported by the forward chaining time-series cross validation is then the average of the values computed in the iterations. This method allows us to obtain a more generalizable model by aggregating metrics over several folds on the temporal order of the data.
	
	\begin{figure*}[!htb]
		\centering
		\caption{Scenario 1: prediction comparison.}
		\pgfplotsset{my style/.append style={axis x line=left, axis y line=left}}
\definecolor{spirodiscoball}{rgb}{0.06, 0.75, 0.99}
\definecolor{asparagus}{rgb}{0.53, 0.66, 0.42}
\definecolor{blue-violet}{rgb}{0.54, 0.17, 0.89}
\definecolor{bblack}{rgb}{0.5, 0.5, 0.5}
\definecolor{frenchrose}{rgb}{0.96, 0.29, 0.54}
\begin{tikzpicture}
	\small
	\begin{axis} 
		[ 
		ylabel = {Quarterly real GDP (billions of Ariary)},
		scale=0.85,
		legend columns=-1,
		legend style={/tikz/every even column/.append style={column sep=0.5cm}},
		legend style={at={(0.5,-0.325)},anchor=north},
		xtick={10,11,12,13,14,15,16,17,18,19,20,21,22,23,24,25}, 
		ytick={3250,4000,...,6000},
		xticklabels={2019Q1, 2019Q2, 2019Q3, 2019Q4, 2020Q1, 2020Q2, 2020Q3, 2020Q4, 2021Q1, 2021Q2, 2021Q3, 2021Q4, 2022Q1, 2022Q2, 2022Q3, 2022Q4},
		xtick align=center,
		ytick align=center,
		xtick pos=bottom,
		ytick pos=left,
		x tick label style={rotate=90, anchor=east},
		x tick label style={/pgf/number format/.cd,%
			scaled x ticks = false,
			set thousands separator={},
			fixed},
		x=1cm,
		xmin=9, xmax=26,
		ymin=3000, ymax=6500]  
		
		\addplot[
		color=bblack, very thick, mark=o,
		]
		coordinates {
			(10,5169.36588724859)
			(11,5298.09950096296)
			(12,5544.56134859134)
			(13,5868.90767732018)
			(14,5535.74840734316)
			(15,4794.10222966737)
			(16,4989.91140193162)
			(17,4999.38312900457)
			(18,4719.16649272635)
			(19,5411.41790601372)
			(20,5558.165723963)
			(21,5796.63589901305)
			(22,5266.58989753453)
			(23,5429.08368221394)
			(24,5559.64351676758)
			(25,5921.88042491598)
		};
		\addlegendentry{Actual}
		
		\addplot[
		mark options={solid, color=asparagus}, color=asparagus, mark = diamond*, densely dashed, thick
		]
		coordinates {
			(10,5048.96497151207)
			(11,5359.32961953213)
			(12,5431.22871198545)
			(13,5535.85036459462)
			(14,5126.15851516614)
			(15,5507.65017537195)
			(16,5464.84886907848)
			(17,5801.31236307633)
			(18,5257.72755470987)
			(19,5638.53511819276)
			(20,5645.40299441381)
			(21,5865.94963308446)
			(22,5381.30046127546)
			(23,5763.0889176886)
			(24,5749.60286300396)
			(25,6040.47358029721)
		};
		\addlegendentry{AR(4)}
		
		\addplot[
		mark options={solid, color=blue-violet}, color=blue-violet, mark = square*, thick, densely dashed
		]
		coordinates {
			(10,5201.97810837391)
			(11,5206.4197645443)
			(12,5458.57253738996)
			(13,5651.02398749393)
			(14,5356.10837585171)
			(15,3359.32000135234)
			(16,3461.43068666571)
			(17,4997.68356449173)
			(18,4905.77313568732)
			(19,4665.85489277403)
			(20,5168.54443011946)
			(21,5607.55331069745)
			(22,5117.56315084452)
			(23,5375.9237285272)
			(24,5550.9830855143)
			(25,5793.85163519427)
		};
		\addlegendentry{OLS-log}
		
		\addplot[mark options={solid, color=spirodiscoball}, color=spirodiscoball, mark =10-pointed star, thick, dashed
		]
		coordinates {
			(10,5235.48092579623)
			(11,5242.85591462685)
			(12,5517.49606057464)
			(13,5801.88043954622)
			(14,5322.56344115076)
			(15,4693.64580604338)
			(16,4867.09395720601)
			(17,5437.82138423128)
			(18,5105.49416387839)
			(19,5286.32615400253)
			(20,5541.40452916133)
			(21,6141.20239012574)
			(22,5522.75545854144)
			(23,5579.86089631785)
			(24,5741.44916901832)
			(25,6035.33532754766)
		};
		\addlegendentry{OLS-RS}
		
		\addplot[
		mark options={solid, color=frenchrose}, color=frenchrose, mark = star, very thick 
		]
		coordinates {
			(10,5151.92083848079)
			(11,5205.5395830996)
			(12,5452.26411182528)
			(13,5650.55029781861)
			(14,5248.56052443518)
			(15,4805.1495481521)
			(16,4941.24062082135)
			(17,5377.07123562804)
			(18,5120.55982682205)
			(19,5265.99448462916)
			(20,5437.05194315869)
			(21,5840.51022961236)
			(22,5433.08857905918)
			(23,5477.53593079435)
			(24,5580.04830422416)
			(25,5839.90141191688)
		};
		\addlegendentry{Ensemble Model}
	\end{axis}
\end{tikzpicture}
		\label{fig:scn1pred}
	\end{figure*}
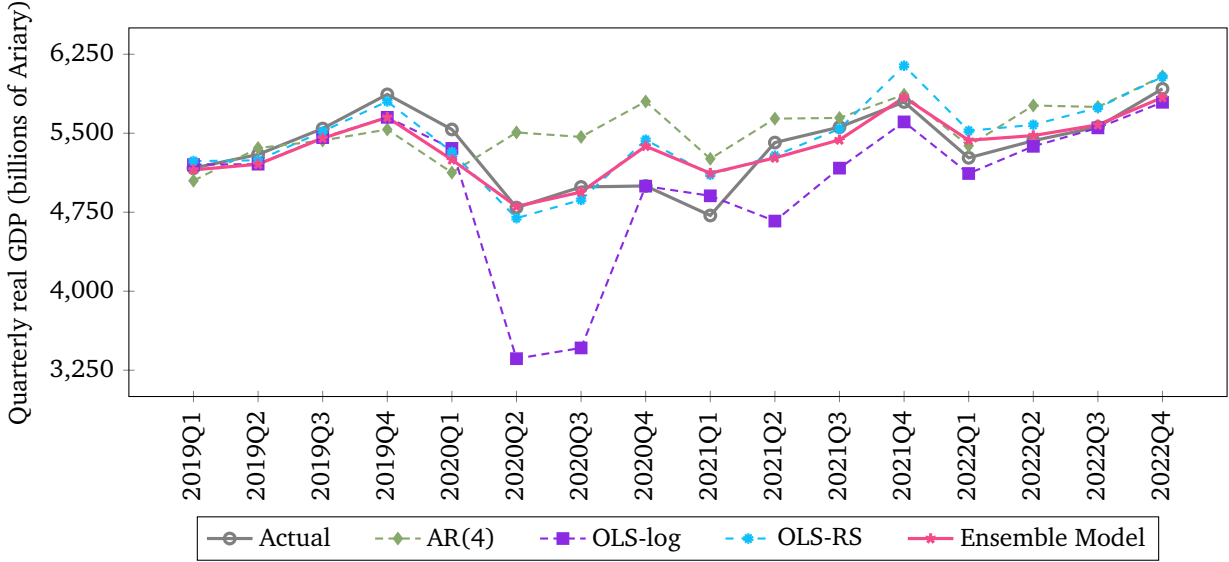
	
	\subsection{Models performance evaluation}
	To assess the predictive performance of the proposed machine learning models, we used the root mean square error (RMSE), mean average error (MAE), and mean absolute percentage error (MAPE):	
	\begin{align*}
		\mathit{RMSE} &= \sqrt{\dfrac{1}{N}\displaystyle\sum_{i=1}^{N}(y_i- \widehat{y}_i)^2},\\
		\mathit{MAE}  &= \dfrac{1}{N} \sum_{i=1}^{N}\left|y_i- \widehat{y}_i\right|,\\
		\mathit{MAPE} &= \dfrac{1}{N} \sum_{i=1}^{N}\left|\dfrac{y_i - \widehat{y}_i}{y_i}\right|\times 100,
	\end{align*}
	where $\widehat{y}_i$ denotes the predicted value.

	\section{Empirical results}
	\subsection{Scenario 1}\label{subsec:scn1}
	The train and test results are shown in \Tabs{tab:scn1TrainTest}. During the training phase, RFR emerged as the dominant model, exhibiting superior performance across all metrics (RMSE, MAE, MAPE).  However, transitioning to the testing phase reveals a shift in model effectiveness. Here, XGBoost diplays better performance in terms of RMSE, showcasing its superior generalization ability to unseen data.  Additionally, E-Net demonstrates robust performance, particularly in terms of MAE and MAPE.
	
	\begin{table}[!htb]
		\centering
		\caption{Scenario 1: performance metrics comparison.}
		\begin{tabular}{@{}lD{.}{.}{4.2}D{.}{.}{5.2}D{.}{.}{2.1}@{}}
			\toprule
			\multicolumn{4}{c}{\textbf{Train metrics}} \\
			\midrule
			Models      & \multicolumn{1}{c}{RMSE} & \multicolumn{1}{c}{MAE} & \multicolumn{1}{r}{MAPE} \\
			\midrule
			Ridge & 131.281 & 105.473 & 2.334 \\
			Lasso & 127.340 & 99.862 & 2.217 \\
			E-Net & 131.473 & 105.763 & 2.340 \\
			PCR   & 132.580 & 106.766 & 2.369 \\
			RFR & \grasi{64.545} & \gras{49.640} & \grasiii{1.093} \\
			$k$-NN  & 140.260 & 112.928 & 2.486 \\
			SVR   & 131.231 & 94.210 & 2.084 \\
			XGBoost & 128.645 & 101.302 & 2.246 \\
			\midrule
			\multicolumn{4}{c}{\textbf{Test metrics}} \\
			\midrule
			& \multicolumn{1}{c}{RMSE} & \multicolumn{1}{c}{MAE} & \multicolumn{1}{r}{MAPE} \\
			\midrule
			Ridge & 197.326 & 152.123 & 2.784 \\
			Lasso & 201.799 & 161.440 & 2.956 \\
			E-Net & 196.942 &  \gras{151.774} & \grasii{2.778} \\
			PCR   & 331.073 & 256.513 & 4.482 \\
			RFR & 314.171 & 273.184 & 5.254 \\
			$k$-NN  & 333.332 & 301.597 & 5.802 \\
			SVR   & 264.039 & 195.593 & 3.475 \\
			XGBoost & \grasi{179.402} & 154.456 & 2.927 \\
			\multicolumn{4}{c}{\textit{Ensemble Model vs.~Benchmark models}} \\
			Ensemble Model & \grasi{184.614} & \gras{144.879} & \grasiii{2.761} \\
			{AR(4)} & 410.010 & 294.220 & 5.698 \\
			{OLS-log} & 576.151 & 338.907 & 6.600 \\
			{OLS-RS} & 208.219 & 166.581 & 3.169 \\
			\bottomrule
		\end{tabular}%
		\label{tab:scn1TrainTest}%
	\end{table}%
	
	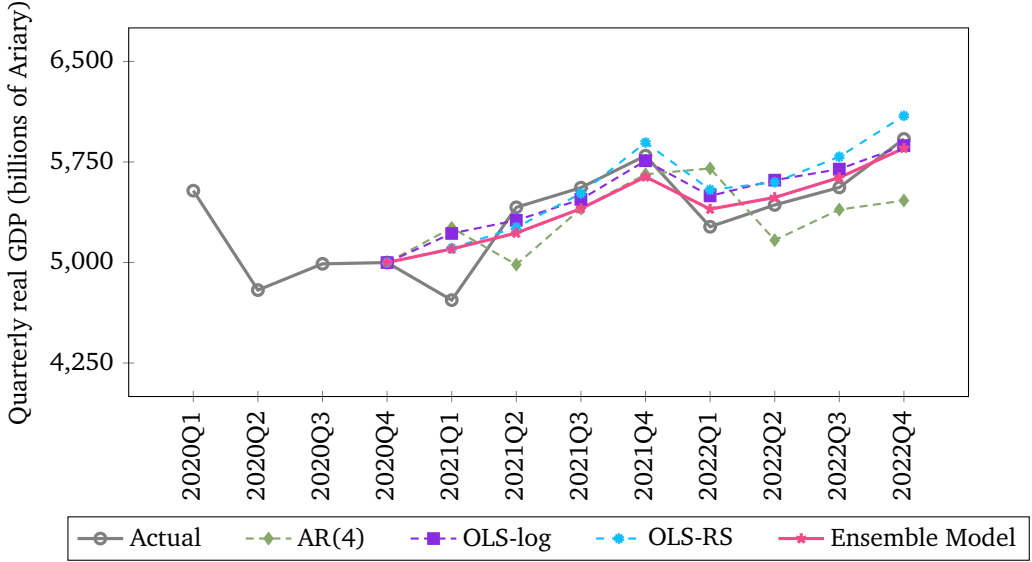
\begin{figure*}[!htb]
		\centering
		\caption{Scenario 2: prediction comparison.}	
		\pgfplotsset{my style/.append style={axis x line=left, axis y line=left}}

\definecolor{spirodiscoball}{rgb}{0.06, 0.75, 0.99}
\definecolor{asparagus}{rgb}{0.53, 0.66, 0.42}
\definecolor{blue-violet}{rgb}{0.54, 0.17, 0.89}
\definecolor{bblack}{rgb}{0.5, 0.5, 0.5}
\definecolor{frenchrose}{rgb}{0.96, 0.29, 0.54}
\begin{tikzpicture}
	\small
	\begin{axis} 
		[
		xtick align=center,
		ytick align=center,
		xtick pos=bottom,
		ytick pos=left,
		ylabel = {Quarterly real GDP (billions of Ariary)},
		scale=0.85,
		legend columns=-1,
		legend style={/tikz/every even column/.append style={column sep=0.5cm}},
		legend style={at={(0.5,-0.325)},anchor=north},
		xtick={10,11,12,13,14,15,16,17,18,19,20,21}, 
		ytick={4250,5000,...,6250},
		xticklabels={2020Q1, 2020Q2, 2020Q3, 2020Q4, 2021Q1, 2021Q2, 2021Q3, 2021Q4, 2022Q1, 2022Q2, 2022Q3, 2022Q4},
		x tick label style={rotate=90, anchor=east},
		x tick label style={/pgf/number format/.cd,%
			scaled x ticks = false,
			set thousands separator={},
			fixed},
		x=1cm,
		xmin=9, xmax=22,
		ymin=4000, ymax=6750]  
		
		\addplot[
		color=bblack, very thick,mark=o,
		]
		coordinates {
			(10,5535.74840734316)
			(11,4794.10222966737)
			(12,4989.91140193162)
			(13,4999.38312900457)
			(14,4719.16649272635)
			(15,5411.41790601372)
			(16,5558.165723963)
			(17,5796.63589901305)
			(18,5266.58989753453)
			(19,5429.08368221394)
			(20,5559.64351676758)
			(21,5921.88042491598)
		};
		\addlegendentry{Actual}
		
		\addplot[
		mark options={solid, color=asparagus},color=asparagus, mark = diamond*, densely dashed, thick
		]
		coordinates {
			(13,4999.38312900457)
			(14,5258.32476605534)
			(15,4983.82859259457)
			(16,5402.26517890979)
			(17,5657.83825571245)
			(18,5701.74265642693)
			(19,5166.49222369526)
			(20,5394.70802077455)
			(21,5462.19301034902)
		};
		\addlegendentry{AR(4)}
		\addplot[
		mark options={solid, color=blue-violet},color=blue-violet, mark = square*, thick, densely dashed
		]
		coordinates {
			(13,4999.38312900457)
			(14,5216.45329872584)
			(15,5314.24669318014)
			(16,5470.46978634529)
			(17,5758.95863599134)
			(18,5496.72937566074)
			(19,5614.31107695536)
			(20,5696.7145945681)
			(21,5871.92754657226)
		};
		\addlegendentry{OLS-log}
		\addplot[
		mark options={solid, color=spirodiscoball},color=spirodiscoball, mark =10-pointed star, thick, dashed
		]
		coordinates {
			(13,4999.38312900457)
			(14,5105.14122549294)
			(15,5261.98383845791)
			(16,5514.75766580287)
			(17,5894.84907776771)
			(18,5542.27938079752)
			(19,5598.48214517136)
			(20,5788.77420692978)
			(21,6094.12900332923)
		};
		\addlegendentry{OLS-RS}
		
		\addplot[
		mark options={solid, color=frenchrose},color=frenchrose, mark = star, very thick, 
		]
		coordinates {
			(13,4999.38312900457)
			(14,5100.82566199919)
			(15,5219.49734979029)
			(16,5401.88767221788)
			(17,5639.2807557904)
			(18,5398.14943480633)
			(19,5484.83675926925)
			(20,5633.23722979437)
			(21,5852.16182679476)
		};
		\addlegendentry{Ensemble Model}
	\end{axis}
\end{tikzpicture}
		\label{fig:scn2pred}
	\end{figure*}
	
	The Ensemble Model reveals competitive performance across RMSE, MAE, and MAPE with respect to the benchmark models. This highlights the strength of combining multiple machine learning models, as evidenced by its effective performance compared to individual algorithms. The Ensemble Model even outperforms all individual machine learning models in terms of MAE and MAPE during the testing phase. Overall, the results indicate that the Ensemble Model, along with XGBoost and E-Net and Ridge, offers a balanced and accurate forecasting solution for GDP in the given economic context
	
	\Figs{fig:scn1pred} displays the comparison the prediction of the Ensemble Model and the benchmark models. All models performed relatively well, but OLS-log displays the largest discrepancy in 2020Q2 and 2020Q3, possibly due to its sensitivity to outliers.  The economic disruptions caused by the COVID-19 pandemic might have introduced outliers that OLS-log struggled to capture. We note that the OLS-RS model, which incorporates robust scaling, demonstrates superior performance compared to the standard OLS model. This emphasizes the effectiveness of robust scaling in handling outliers contributing to improved model performance. It suggests that, in this specific context, OLS-RS model with robust scaling is a preferable choice over OLS-log model.
	
	\begin{table}[!htb]
		\centering
		\caption{Scenario 2: performance metrics comparison.}
		\begin{tabular}{@{}lD{.}{.}{4.2}D{.}{.}{5.2}D{.}{.}{2.1}@{}}
			\toprule
			\multicolumn{4}{c}{\textbf{Train metrics}} \\
			\midrule
			Models      & \multicolumn{1}{c}{RMSE} & \multicolumn{1}{c}{MAE} & \multicolumn{1}{r}{MAPE} \\
			\midrule
			Ridge & 128.969 & 103.590 & 2.253 \\
			Lasso & 127.062 & 101.012 & 2.204 \\
			E-Net & 129.176 & 103.858 & 2.259 \\
			PCR   & 140.159 & 111.943 & 2.434 \\
			RFR & \grasi{77.221} & \gras{58.199} & \grasiii{1.248} \\
			$k$-NN  & 138.108 & 107.789 & 2.342 \\
			SVR   & 132.914 & 99.132 & 2.136 \\
			XGBoost & 140.223 & 115.134 & 0.025 \\
			\midrule
			\multicolumn{4}{c}{\textbf{Test metrics}} \\
			\midrule
			Models    & \multicolumn{1}{c}{RMSE} & \multicolumn{1}{c}{MAE} & \multicolumn{1}{r}{MAPE} \\
			\midrule
			Ridge & 201.610 & 169.001 & 3.116 \\
			Lasso & 210.394 & 189.657 & 3.439 \\
			E-Net & 200.970 & 167.943 & 3.099 \\
			PCR   & 216.242 & 182.616 & 3.384 \\
			RFR & 251.587 & 214.974 & 4.025 \\
			$k$-NN  & 422.254 & 392.298 & 7.670 \\
			SVR   & 226.896 & 182.811 & 3.335 \\
			XGBoost & \grasi{140.701} & \gras{113.067} & \grasii{2.134} \\
			\multicolumn{4}{c}{\textit{Ensemble Model vs.~Benchmark models}} \\
			Ensemble Model & \grasi{184.614} & \gras{144.879} & \grasii{2.761} \\
			{AR(4)} & 356.006 & 322.977 & 6.044 \\
			{OLS-log} & 216.334 & 165.278 & 3.206 \\
			{OLS-RS} & 216.143 & 211.617 & 3.846 \\
			\bottomrule
		\end{tabular}%
		\label{tab:scn2TrainTest}%
	\end{table}%
	
	\begin{figure*}[!ht]
		\centering
		\caption{Scenario 3: prediction comparison.}
		\pgfplotsset{my style/.append style={axis x line=left, axis y line=left}}
\definecolor{spirodiscoball}{rgb}{0.06, 0.75, 0.99}
\definecolor{asparagus}{rgb}{0.53, 0.66, 0.42}
\definecolor{blue-violet}{rgb}{0.54, 0.17, 0.89}
\definecolor{bblack}{rgb}{0.5, 0.5, 0.5}
\definecolor{frenchrose}{rgb}{0.96, 0.29, 0.54}
\begin{tikzpicture}
	\small
	\begin{axis} 
		[  		            
		xtick align=center,
		ytick align=center,
		xtick pos=bottom,
		ytick pos=left,
		ylabel = {Quarterly real GDP (billions of Ariary)},
		scale=0.85,
		legend columns=-1,
		legend style={/tikz/every even column/.append style={column sep=0.5cm}},
		legend style={at={(0.5,-0.325)},anchor=north},
		xtick={10,11,12,13,14,15,16,17}, 
		ytick={4250,5000,...,6250},
		xticklabels={2021Q1, 2021Q2, 2021Q3, 2021Q4, 2022Q1, 2022Q2, 2022Q3, 2022Q4},
		x tick label style={rotate=90, anchor=east},
		x tick label style={/pgf/number format/.cd,%
			scaled x ticks = false,
			set thousands separator={},
			fixed},
		x=1cm,
		xmin=9, xmax=18,
		ymin=4000, ymax=6600]  
		
		\addplot[
		color=bblack, very thick,mark=o,
		]
		coordinates {
			(10,4719.16649272635)
			(11,5411.41790601372)
			(12,5558.165723963)
			(13,5796.63589901305)
			(14,5266.58989753453)
			(15,5429.08368221394)
			(16,5559.64351676758)
			(17,5921.88042491598)
		};
		\addlegendentry{Actual}
		
		\addplot[
		mark options={solid, color=asparagus},color=asparagus, mark = diamond*, densely dashed, thick
		]
		coordinates {
			(13,5796.63589901305)
			(14,5503.44237759316)
			(15,5413.81163033866)
			(16,5443.97613379531)
			(17,5406.00076948837)
		};
		\addlegendentry{AR(4)}
		\addplot[
		mark options={solid, color=blue-violet},color=blue-violet, mark = square*, thick, densely dashed
		]
		coordinates {
			(13,5796.63589901305)
			(14,5461.26671710502)
			(15,5580.18118787998)
			(16,5659.12700011228)
			(17,5814.12552746458)
		};
		\addlegendentry{OLS-log}
		\addplot[
		mark options={solid, color=spirodiscoball},color=spirodiscoball, mark =10-pointed star, thick, dashed
		]	
		coordinates {
			(13,5796.63589901305)
			(14,5508.04940414637)
			(15,5580.84039788483)
			(16,5780.15154488002)
			(17,6095.55851040211)
		};
		\addlegendentry{OLS-RS}
		
		\addplot[
		mark options={solid, color=frenchrose},color=frenchrose, mark = star, very thick, 
		]
		coordinates {
			(13,5796.63589901305)
			(14,5432.89547601625)
			(15,5500.26011278476)
			(16,5642.57277722352)
			(17,5871.28736718424)
		};
		\addlegendentry{Ensemble Model}
	\end{axis}
\end{tikzpicture}
		\label{fig:scn3pred}
	\end{figure*}
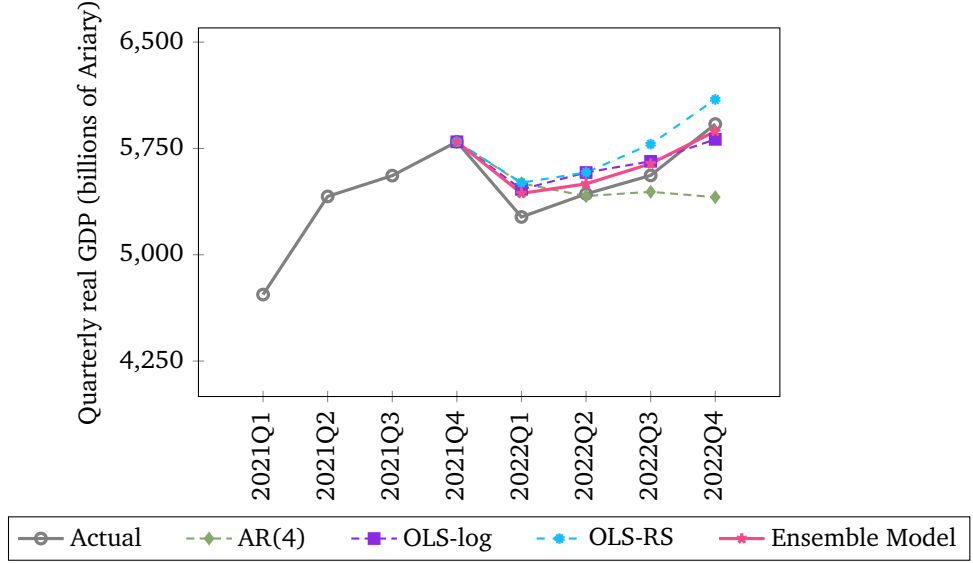

	\subsection{Scenario 2}\label{subsec:scn2}
	This scenario enables us to assess the overall impact of adding the 2019-2020 data. As shown in \Tabs{tab:scn2TrainTest} in the train metrics panel, this led to slightly higher RMSE, MAE and MAPE values for some models compared with \Scn{subsec:scn1}{1}, suggesting potential difficulty in capturing the unique effects of the COVID-19 pandemic. In \Scn{subsec:scn2}{2}, RFR exhibits the best performance in terms of RMSE, MAE, and MAPE. However, during the testing phase, XGBoost performs the best, indicating its adaptability to unseen data. Conversely, $k$-NN shows a notable decline in performance during testing, indicating challenges in adapting to the addition of COVID-19 data in the training set. This decline could be attributed to the algorithm perceiving similarities among indicators due to policy measures taken in response to the COVID-19 crisis (e.g., VAT, government expenditure \cite{INSTAT20221, MEF2023}). During training, these indicators may have exhibited patterns similar to non-crisis periods, leading the algorithm to associate them with normal situations. Therefore, in the testing phase, the attributes in the neighborhood in this case includes outliers that impact the predictive accuracy of $k$-NN. 
	
	The Ensemble Model consistently outperforms the benchmark models across all metrics in the testing phase. Furthermore, when comparing the Ensemble Model to individual machine learning models, we note that it positions itself between XGBoost and the other models in terms of RMSE, MAE, and MAPE. 
	
	As shown in \Tabs{tab:scn2TrainTest} and \Figs{fig:scn2pred}, OLS-log generally performs slightly better than OLS-RS on the test set, whereas AR(4) performs the worst. When we compare AR(4) to \Scn{subsec:scn1}{1} in the 2021Q1 to 2022Q4 period, we note that the model struggles to capture the abrupt changes and variations induced by the pandemic.
	
	\begin{table}[!htb]
		\centering
		\caption{Scenario 3: performance metrics comparison.}
		\begin{tabular}{@{}lD{.}{.}{4.2}D{.}{.}{5.2}D{.}{.}{2.1}@{}}
			\toprule
			\multicolumn{4}{c}{\textbf{Train metrics}} \\
			\midrule
			Models & \multicolumn{1}{c}{RMSE} & \multicolumn{1}{c}{MAE} & \multicolumn{1}{r}{MAPE} \\
			\midrule
			Ridge & 134.890 & 108.156 & 2.327 \\
			Lasso & 132.590 & 106.655 & 2.302 \\
			E-Net & 135.424 & 108.660 & 2.337 \\
			PCR   & 143.032 & 116.207 & 2.492 \\
			RFR & \grasi{77.259} & \gras{57.787} & \grasiii{1.228} \\
			$k$-NN  & 184.228 & 137.044 & 2.875 \\
			SVR   & 144.224 & 102.886 & 2.183 \\
			XGBoost & 139.713 & 114.601 & 2.434 \\
			\midrule
			\multicolumn{4}{c}{\textbf{Test metrics}} \\
			\midrule
			Models    & \multicolumn{1}{c}{RMSE} & \multicolumn{1}{c}{MAE} & \multicolumn{1}{r}{MAPE} \\
			\midrule
			Ridge & 161.325 & 157.303 & 2.781 \\
			Lasso & 200.248 & 196.889 & 3.442 \\
			E-Net & 157.515 & 153.301 & 2.713 \\
			PCR & 162.814 & 154.534 & 2.741 \\
			RFR & 186.641 & 131.574 & 2.383 \\
			$k$-NN & 327.164 & 304.312 & 5.796 \\
			SVR & 211.216 & 197.585 & 3.481 \\
			XGBoost & \grasi{90.119} & \gras{72.893} & \grasii{1.322} \\
			\multicolumn{4}{c}{\textit{Ensemble Model vs.~Benchmark models}} \\
			Ensemble Model & \grasi{109.065} & \gras{90.823} & \grasii{1.685} \\
			AR(4) & 289.760 & 220.918 & 3.893 \\
			OLS-log   & 143.385 & 138.253 & 2.522 \\
			OLS-RS  & 200.075 & 196.851 & 3.570 \\
			\bottomrule
		\end{tabular}%
		\label{tab:scn3TrainTest}%
	\end{table}%
	
	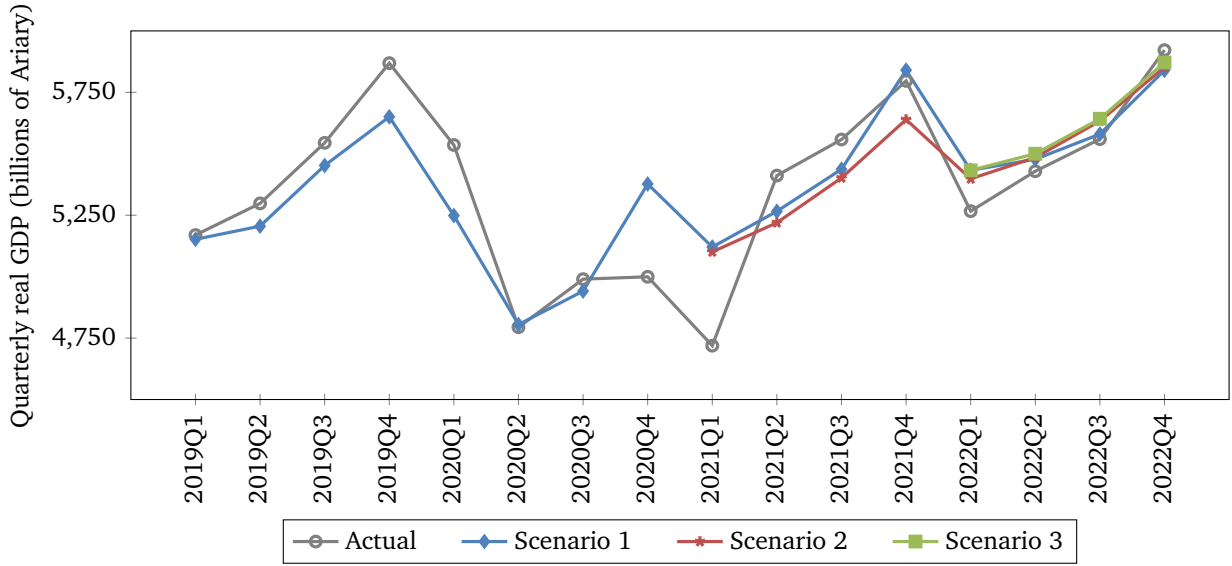
\begin{figure*}[!htb]
		\centering
		\caption{Ensemble Model prediction comparison across Scenario 1,2,3}
		\pgfplotsset{my style/.append style={axis x line=left, axis y line=left}}
\definecolor{bblack}{rgb}{0.5, 0.5, 0.5}
\definecolor{bblue}{HTML}{4F81BD}
\definecolor{rred}{HTML}{C0504D}
\definecolor{ggreen}{HTML}{9BBB59}

\begin{tikzpicture}
	\small
	\begin{axis} 
		[
		ylabel = { Quarterly real GDP (billions of Ariary)},        		           
		xtick align=center,
		ytick align=center,
		xtick pos=bottom,
		ytick pos=left,
		scale=0.85,
		legend columns=-1,
		legend style={/tikz/every even column/.append style={column sep=0.5cm}},
		legend style={at={(0.5,-0.325)},anchor=north},
		xtick={10,11,12,13,14,15,16,17,18,19,20,21,22,23,24,25}, 
		ytick={4750,5250,...,5750}, 
		xticklabels={2019Q1, 2019Q2, 2019Q3, 2019Q4, 2020Q1, 2020Q2, 2020Q3, 2020Q4, 2021Q1, 2021Q2, 2021Q3, 2021Q4, 2022Q1, 2022Q2, 2022Q3, 2022Q4},
		x tick label style={rotate=90, anchor=east},
		x tick label style={/pgf/number format/.cd,%
			scaled x ticks = false,
			set thousands separator={},
			fixed},
		x=1cm,
		xmin=9, xmax=26,
		ymin=4500, ymax=6000]  
		
		\addplot[
		color=bblack, very thick, mark=o,
		]
		coordinates {
			(10,5169.36588724859)
			(11,5298.09950096296)
			(12,5544.56134859134)
			(13,5868.90767732018)
			(14,5535.74840734316)
			(15,4794.10222966737)
			(16,4989.91140193162)
			(17,4999.38312900457)
			(18,4719.16649272635)
			(19,5411.41790601372)
			(20,5558.165723963)
			(21,5796.63589901305)
			(22,5266.58989753453)
			(23,5429.08368221394)
			(24,5559.64351676758)
			(25,5921.88042491598)
		};
		\addlegendentry{Actual}
		
		\addplot[
		mark options={solid, color=bblue}, color=bblue, mark = diamond*, very thick, 
		]
		coordinates {
			(10,5151.92083848079)
			(11,5205.5395830996)
			(12,5452.26411182528)
			(13,5650.55029781861)
			(14,5248.56052443518)
			(15,4805.1495481521)
			(16,4941.24062082135)
			(17,5377.07123562804)
			(18,5120.55982682205)
			(19,5265.99448462916)
			(20,5437.05194315869)
			(21,5840.51022961236)
			(22,5433.08857905918)
			(23,5477.53593079435)
			(24,5580.04830422416)
			(25,5839.90141191688)
		};
		\addlegendentry{Scenario 1}
		
		\addplot[
		color=rred, mark = star, very thick,
		]
		coordinates {
			(18,5100.82566199919)
			(19,5219.49734979029)
			(20,5401.88767221788)
			(21,5639.2807557904)
			(22,5398.14943480633)
			(23,5484.83675926925)
			(24,5633.23722979437)
			(25,5852.16182679476)
		};
		\addlegendentry{Scenario 2}
		
		\addplot[
		mark options={solid, color=ggreen},color=ggreen, mark = square*, very thick
		]
		coordinates {
			(22,5432.89547601625)
			(23,5500.26011278476)
			(24,5642.57277722352)
			(25,5871.28736718424)
		};
		\addlegendentry{Scenario 3}		
		
	\end{axis}
\end{tikzpicture}
		\label{fig:prediction}
	\end{figure*}
	
	\begin{figure*}[!htb]
		\centering
		\caption{Forecast percentage errors comparison.}
		\definecolor{bblue}{HTML}{4F81BD}
\definecolor{rred}{HTML}{C0504D}
\definecolor{ggreen}{HTML}{9BBB59}

\begin{tikzpicture}
	\small
	\begin{axis}[
		scale=0.85,
		major x tick style = transparent,
		major y tick style = transparent,
		ybar=2*\pgflinewidth,
		bar width=5pt,
		ymajorgrids = true,
		ylabel = {Percentage (\%)},
		x = 1cm,
		x tick label style={rotate=90, anchor=east},
		symbolic x coords={2019Q1, 2019Q2, 2019Q3, 2019Q4, 2020Q1, 2020Q2, 2020Q3, 2020Q4, 2021Q1, 2021Q2, 2021Q3, 2021Q4, 2022Q1, 2022Q2, 2022Q3, 2022Q4},
		xtick = data,
		scaled y ticks = false,
		enlarge x limits=0.05,
		ymin=-5,
		xmin = 2019Q1,
		legend style={
			at={(0.5,-0.35)},
			anchor=north,
			legend columns=-1,
			/tikz/every even column/.append style={column sep=0.5cm}
		}
		]
		\addplot[
		style={bblue,fill=bblue,mark=none}
		]
		coordinates {
			(2019Q1,0.520684828437342)
			(2019Q2,-1.50698640432874)
			(2019Q3,-1.07645376890686)
			(2019Q4,-2.35565038712568)
			(2020Q1,-4.35065482117159)
			(2020Q2,-0.607822893868498)
			(2020Q3,-1.61308627790235)
			(2020Q4,8.01630821283969)
			(2021Q1,8.46182748347142)
			(2021Q2,-2.66305112825219)
			(2021Q3,-1.38649293409628)
			(2021Q4,3.3594794233708)
			(2022Q1,3.9075014495297)
			(2022Q2,1.82817523295499)
			(2022Q3,1.99999109873506)
			(2022Q4,0.517640167608604)
		};
		
		\addplot[
		style={rred,fill=rred,mark=none}
		]
		coordinates {
			(2021Q1,7.68887372840534)
			(2021Q2,-3.13935529530023)
			(2021Q3,-2.34384946320878)
			(2021Q4,-0.369002825448897)
			(2022Q1,3.17846921172003)
			(2022Q2,1.52925703741517)
			(2022Q3,2.17798929314841)
			(2022Q4,0.21501759524944)
		};
		
		\addplot[
		style={ggreen,fill=ggreen,mark=none}
		]
		coordinates {
			(2022Q1,3.09331790001902)
			(2022Q2,1.48445471338549)
			(2022Q3,1.92515700677167)
			(2022Q4,-0.02020511814595)
		};
		
		\legend{Scenario 1, Scenario 2, Scenario 3}
		
	\end{axis}
\end{tikzpicture}
		\label{fig:QPE}
	\end{figure*}
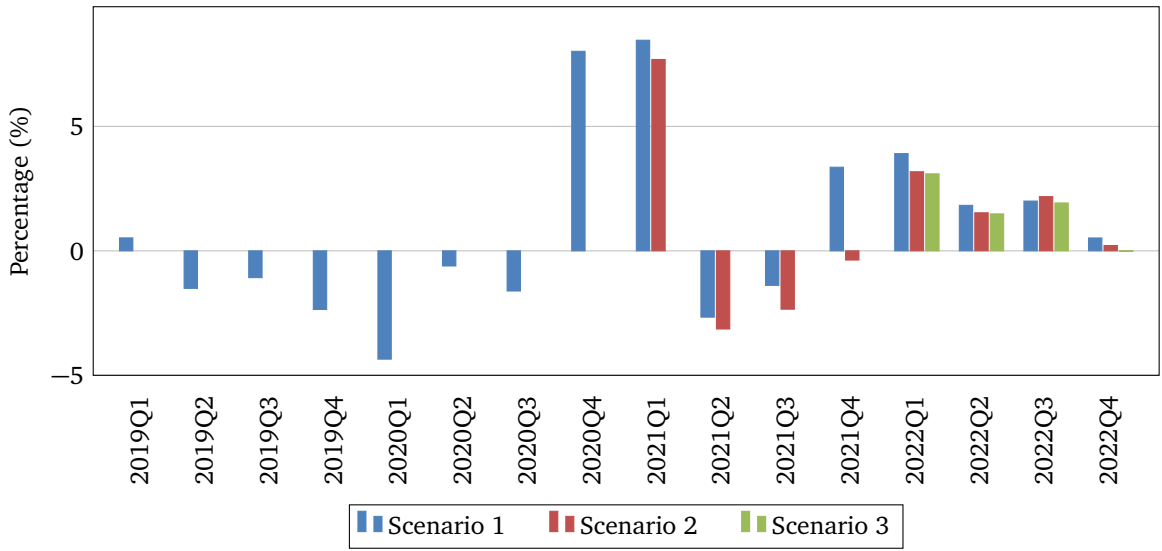

	\subsection{Scenario 3}\label{subsec:scn3}
	In this scenario, the Ensemble Model outperforms all other models on both train and test sets across all metrics (RMSE, MAE, MAPE), indicating its superior accuracy and generalization ability (see \Tabs{tab:scn3TrainTest}). RFR performs exceptionally well on the train set due to its flexibility in capturing complex relationships, but this might lead to overfitting on the training data as evident by its poorer performance on the test set. XGBoost achieves a good balance between train and test set performance, suggesting it learns from the data effectively without overfitting. Ridge, Lasso, and E-Net perform decently on the train set but are significantly outperformed by RFR and XGBoost, highlighting the potential benefits of more complex modeling approaches for this specific scenario. $k$-NN and SVR underperform most other models.
	
	While econometric models generally outperform the AR(4) benchmark, OLS-log performs the best on the test set. OLS-RS also follows closely the  trajectory of the actual value, though not as much as OLS-log. Notably, AR(4), limited by its dependence on past data, particularly struggles to predict accurately in 2022Q3 and 2022Q4.

	\section{Discussion}
	We first combine the forecasts for the three scenarios in \Figs{fig:prediction}. We also report in \Figs{fig:QPE}  the quarterly forecast percentage errors of each model for the test period of 2019Q1 to 2022Q4. Both figures show that larger deviation can be observed in 2020Q4 and 2021Q1.
	
	In \Scn{subsec:scn1}{1}, where the model is trained up to 2018Q4 and tested on the period from 2019Q1 to 2022Q4, it serves as a baseline for evaluating the model's predictive performance. This scenario captures the model's ability to generalize to the economic conditions of the early 2020 based on prior knowledge. During the stable period of 2019, the Ensemble Model exhibited close alignment between its predictions and the actual values. However, as the Covid-19 crisis unfolded in 2020 and early 2021, substantial GDP volatility emerged, markedly diverging from the stable trajectory of the previous year. Notably, the Ensemble Model effectively tracked the actual GDP trajectory amidst this volatility, closely mirroring the fluctuations. Subsequently, in the latter half of 2021 and throughout 2022, all models' forecasts converged closely with the actual GDP values, following a similar trajectory.
	
	In \Scn{subsec:scn2}{2}, the training dataset is extended to include observations up to 2020Q4, allowing the model to incorporate information from the initial phases of the COVID-19 pandemic. As shown in \Figs{fig:QPE}, the forecast is improved in 2021Q1 but deviated more from the actual observation in 2021Q2 and 2022Q3. In fact, the model provides an underestimation of GDP during these last two quarters because the model, having seen the economic impact of the pandemic, adjusts its forecasts to better align with the observed outcomes during these quarters.	Consequently, adding observations up to 2020Q4 helps the model better understand the initial effects of the pandemic, leading to a correction in predictions for subsequent quarters.
	
	In \Scn{subsec:scn3}{3}, the training set is further extended to include observations up to 2021Q4. This scenario demonstrates the importance of incorporating the most recent data for accurate predictions. The model in \Scn{subsec:scn3}{3} outperforms both Scenarios \hyperref[subsec:scn1]{1} and \hyperref[subsec:scn1]{2}, showcasing its enhanced adaptability to ongoing economic changes. Therefore, the addition of more observations provides a correction to the trajectory of forecasts. Overestimation tendencies are mitigated as the model learns from more recent economic experiences, showcasing the importance of regularly updating training data for accurate nowcasting.

	\section{Conclusion}
	We evaluated the predictive performance of 8 machine learning algorithms to nowcast Madagascar's real GDP. We trained each algorithm using 10 predictor variables ranging from 2007Q7 to 2022Q4, which consist of quarterly macroeconomic indicators issued by Malagasy authorities. We compared the real-time performance nowcast of each algorithm by examining the RMSE, MAE, and MAPE.  We found that the Ensemble Model obtained as a weighted combination of the nowcast predictions is able to produce accurate estimates of quarterly real GDP compared to the AR(4) and OLS benchmarks. The ensemble model's strength lies in its ability to discern turning points amidst volatile economic conditions, making it well suited for capturing abrupt shifts in activity like those experienced during the COVID-19 period.	Our results also demonstrate that robust scaling for data preprocessing significantly enhances the prediction accuracy. This improvement is evident when we compare two model variants: one using natural logarithm transformation and the other incorporating robust scaling. In addition, by incorporating a more extensive temporal scope in our scenarios, we observed a notable improvement in the precision of our forecasts. This emphasizes the dynamic nature of economic conditions and the importance of taking into account evolving trends in order to make better predictions. Overall, our research not only contributes to the growing literature on nowcasting methodologies, but also offers valuable insights for policymakers and practitioners looking for effective tools for understanding and assessing the current state of the economy.  To the best of our knowledge, our study is one of the first to consider the use of machine learning algorithms to nowcast current Madagascar GDP. Future research could explore incorporating additional data sources and investigating the robustness of these models in different economic scenarios.

\end{document}